\documentclass[pre,aps,twocolumn,amsmath,amssymb,superscriptaddress,floatfix]{revtex4-1}
\usepackage{hyperref}
\usepackage{graphicx}

\begin{document}
\title{Disentangling and modeling interactions in fish with burst-and-coast swimming}

\author{Daniel S. Calovi}
\affiliation{Centre de Recherches sur la Cognition Animale,
Centre de Biologie Int\'egrative (CBI),
Centre  National de la Recherche Scientifique (CNRS) \& Universit\'e de Toulouse (UPS), 31062 Toulouse, France}
\author{Alexandra Litchinko}
\affiliation{Centre de Recherches sur la Cognition Animale,
Centre de Biologie Int\'egrative (CBI),
Centre  National de la Recherche Scientifique (CNRS) \& Universit\'e de Toulouse (UPS), 31062 Toulouse, France}
\author{Valentin Lecheval}
\affiliation{Centre de Recherches sur la Cognition Animale,
Centre de Biologie Int\'egrative (CBI),
Centre  National de la Recherche Scientifique (CNRS) \& Universit\'e de Toulouse (UPS), 31062 Toulouse, France}
\affiliation{Groningen Institute for Evolutionary Life Sciences, University of Groningen, Centre for Life Sciences, Nijenborgh 7, 9747AG Groningen, The Netherlands}
\author{Ugo Lopez}
\affiliation{Centre de Recherches sur la Cognition Animale,
Centre de Biologie Int\'egrative (CBI),
Centre  National de la Recherche Scientifique (CNRS) \& Universit\'e de Toulouse (UPS), 31062 Toulouse, France}
\author{Alfonso P\'erez Escudero}
\affiliation{Department of Physics, Massachusetts Institute of
Technology, Cambridge, MA, USA}
\author{Hugues Chat\'e}
\affiliation{Service de Physique de l'\'Etat Condens\'e, CEA
-- Saclay, 91191 Gif-sur-Yvette, France}
\author{Cl\'ement Sire}
\affiliation{Laboratoire de Physique Th\'eorique, CNRS \& Universit\'e de Toulouse (UPS), 31062
Toulouse, France}
\author{Guy Theraulaz}
\affiliation{Centre de Recherches sur la Cognition Animale,
Centre de Biologie Int\'egrative (CBI),
Centre  National de la Recherche Scientifique (CNRS) \& Universit\'e de Toulouse (UPS), 31062 Toulouse, France}
\email{To whom correspondence
should be addressed. E-mail: guy.theraulaz\@univ-tlse3.fr}

\begin{abstract}
We combine extensive data analyses with a modeling approach to measure, disentangle, and reconstruct the actual functional form of interactions involved in the coordination of swimming in Rummy-nose tetra (\textit{Hemigrammus rhodostomus}). This species of fish performs burst-and-coast swimming behavior that consists of sudden heading changes combined with brief accelerations followed by quasi-passive, straight decelerations. We quantify the spontaneous stochastic behavior of a fish  and the interactions that govern wall avoidance and the attraction and alignment to a neighboring fish, the latter by exploiting general symmetry constraints for the interactions. In contrast with previous experimental works, we find that both attraction and alignment behaviors control the reaction of fish to a neighbor. We then exploit these results to build a model of spontaneous burst-and-coast swimming and interactions of fish, with all parameters being estimated or directly measured from  experiments. This model quantitatively reproduces the key features of the motion and spatial distributions observed in experiments with a single fish and with  two fish. This demonstrates the power of our method that exploits large amounts of data for disentangling and fully characterizing the interactions that govern collective behaviors in animals groups. Moreover, we introduce the notions of ``dumb'' and ``intelligent'' active matter and emphasize and clarify  the strong differences between them.
\end{abstract}

\maketitle

\section{Introduction}

The study of physical or living self-propelled particles -- active matter -- has certainly become a booming field, notably involving biologists and physicists, often working together. Physical examples of active matter include self-propelled Janus colloids \cite{AM1,AM2,AM3,AM4,AM5,AM6,AM7,AM8}, vibrated granular matter \cite{AM9,AM10,AM11}, or self-propulsion mediated by hydrodynamical effects \cite{AM12,AM13},  whereas biological examples are obviously ubiquitous: bacteria, cells, and simply speaking, most animals. In both physical and biological contexts, active matter can organize into rich collective phases. For instance, fish schools can be observed in a disordered swarming phase, or ordered schooling and vortex/milling phases \cite{Tunstrom2013,Calovi2014}.

Yet, there are important difference between ``dumb'' and ``intelligent'' active matter (see the Appendix for a more formal definition and discussion). For the former class, which concerns most physical self-propelled particles, but also, in some context, living active matter, interactions with other particles or obstacles do not modify the intrinsic or ``desired'' velocity of the particles but exert forces whose effect \emph{adds up to its intrinsic velocity}. Intelligent active matter, like fish, birds, or humans, can also interact through physical forces (a human physically pushing an other one or bumping into a wall) but mostly interact through ``social forces''. For instance, a fish or a human wishing to avoid a physical obstacle or an other animal will \emph{modify its intrinsic velocity} in order to never actually touch it. Moreover, a physical force applied to an intelligent active particle, in addition to its direct impact, can elicit a response in the form of a change in its intrinsic velocity (for instance, a human deciding to escape or resist an other human physically pushing her/him). Social forces strongly break the Newtonian law of action and reaction: a fish or a human avoiding a physical obstacle obviously does not exert a contrary force on the obstacle. In addition, even between two animals 1 and 2, the force exerted by 1 on 2 is most often not the opposite of the force exerted by 2 on 1, since social forces commonly depend on stimuli (vision, hearing...) associated to an anisotropic perception: a human will most often react more to another human ahead than behind her/him. Similarly, social forces between two fish or two humans will also depend on their relative velocities or orientations: the need to avoid an other animal will be in general greater when a collision is imminent than if it is unlikely, due to the velocity directions.

Hence, if the understanding of the social interactions that underlie the collective behavior of animal groups is a central question in ethology and behavioral ecology \cite{Camazine2001,Giardina2008}, it has also a clear conceptual interest for physicists, since social and physical forces play very different roles on the dynamics of an active matter particle (see Appendix for details).

These social interactions play a key role in the ability of group members to coordinate their actions and collectively solve a wide range of problems, thus increasing their fitness \cite{Sumpter2010,Krause2002}.
In the past few years, the development of new methods based on machine learning algorithms for automating tracking and behavior analyses of animals in groups has improved to unprecedented levels the precision of available data on social interactions \cite{Branson2009,perez2014,Dell2014}. A wide
variety of biological systems have been investigated using such
methods, from swarms of insects
\cite{Buhl2006,Attanasi2014,Schneider2014} to schools of fish
\cite{Katz2011,Herbert2011,Gautrais2012,Mwaffo2014}, flocks of birds
\cite{Ballerini2008a,Nagy2010,Bialek2014}, groups of mice
\cite{Chaumont2012,Shemesh2013}, herds of ungulates
\cite{Ginelli2015,King2012}, groups of primates
\cite{Strandburg2015,Ballestaa2014}, and human crowds
\cite{Moussaid2011,Gallupa2012}, bringing new insights on behavioral
interactions and their consequences on collective behavior.

The fine-scale analysis of individual-level interactions opens up
new perspectives to develop quantitative and predictive models of
collective behavior. One major challenge is to accurately identify
the contributions and combination of each interaction involved at
individual-level and then to validate with a model their role in the
emergent properties at the collective level
\cite{Lopez2012,Herbert2016}. Several studies on fish schools have
explored ways to infer individual-level interactions directly from
experimental data. The force-map technique \cite{Katz2011} and the
non-parametric inference technique \cite{Herbert2011} have been used
to estimate from experiments involving groups of two fish the
effective turning and speeding forces experienced by an individual.
In the force-map approach, the implicit assumption considers
that fish are particles on which the presence of neighboring fish and physical
obstacles exert ``forces''. Visualizing these effective forces that capture the
coarse-grained regularities of actual interactions has been a first step to
characterize the local individual-level interactions \cite{Katz2011,Herbert2011}. However, none of these
works incorporate nor characterize the intrinsic stochasticity of individual
behavior and neither attempt to validate their findings by building trajectories
from a model.

\begin{figure*}
\centering
\includegraphics[width=\textwidth]{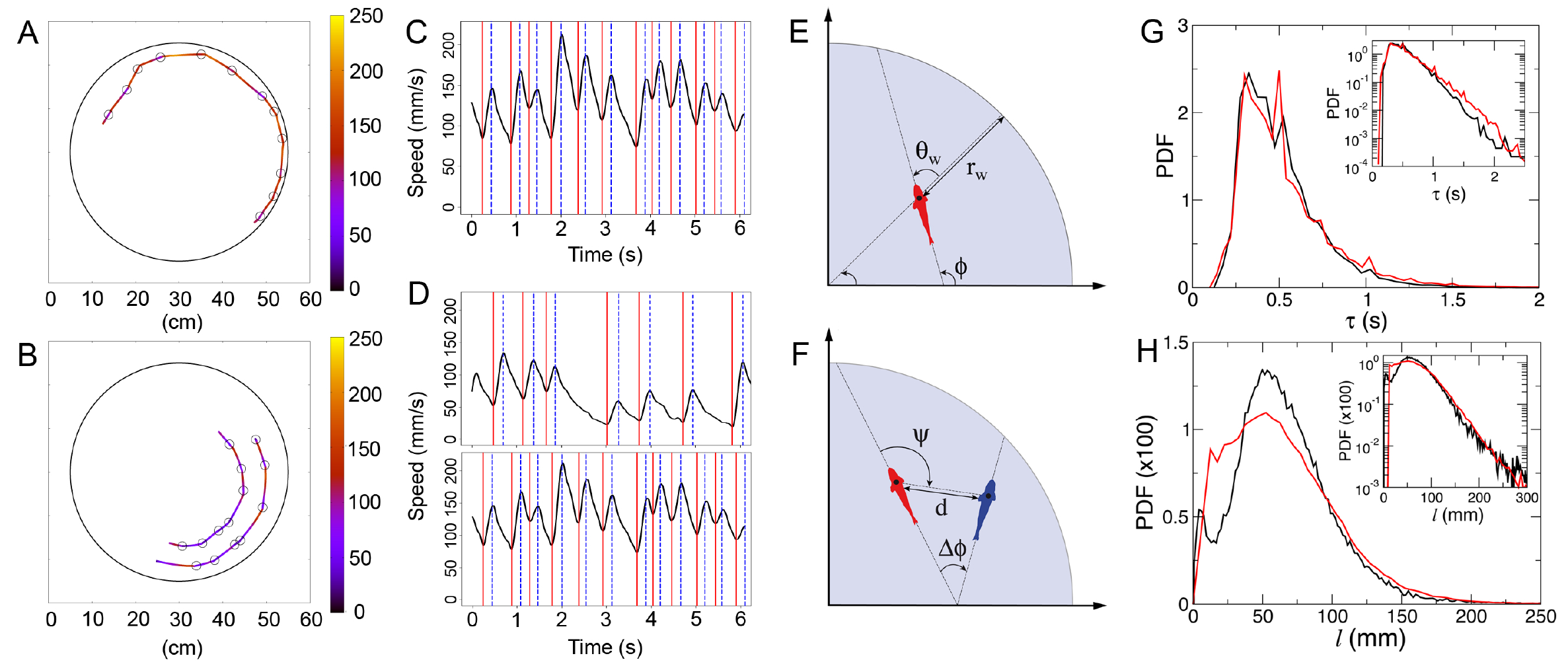}
  \caption{Trajectories along with the bursts (circles) of a fish
    swimming alone (A) and a group of 2 fish (B). The color of
    trajectories indicates instantaneous speed. The corresponding
    speed time series are shown in C and D, along with the
    acceleration/burst phase delimited by red and blue vertical lines. E
    defines the variables $r_{\rm w}$ and $\theta_{\rm w}$  (distance and relative orientation to the wall)
    in order to describe
    the fish interaction with the wall. F defines the relevant
    variables $d$, $\psi$, and $\Delta\phi$ (distance, viewing angle, relative orientation of the focal fish with respect to the other fish)
    in order to describe the
    influence of the blue fish on the red one. G and H show
    respectively the probability distribution function (PDF) of the
    duration and distance traveled between two kicks as measured in the one
(black) and two (red)
    fish experiments (tank of radius $R=250\,$mm). Insets show the corresponding graphs in semi-log scale. \label{Figure1}}
\end{figure*}

On the other hand, only a few models have been developed to connect
a detailed quantitative description of individual-level interactions
with the emergent dynamics observed at a group level
\cite{Herbert2011,Gautrais2012,Mwaffo2014}. The main difficulty to
build such models comes from the entanglement of interactions
between an individual and its physical and social environment. To
overcome this problem, Gautrais et al. \cite{Gautrais2012} have
introduced an incremental approach that consists in first building
from the experiments a model for the spontaneous motion of an
isolated fish \cite{Gautrais2009}. This model is then used as a
dynamical framework to include the effects of interactions of that
fish with the physical environment and with a neighboring fish. The
validation of the model is then based on the agreement of its
predictions with experiments on several observables in different
conditions and group sizes.

In the present work, we use but improve and extend this approach to investigate the swimming behavior and
interactions in the red nose fish \emph{Hemigrammus rhodostomus}. This species performs
a burst-and-coast type of swimming that makes it possible to analyze a
trajectory as a series of discrete behavioral decisions in time and space. This
discreteness of trajectories is exploited to characterize the spontaneous motion
of a fish, to identify the candidate stimuli (\emph{e.g.} the distance, the orientation
and velocity of a neighboring fish, or the distance and orientation of the tank
wall), and to measure their effects on the behavioral response of a fish.  We
assume rather general forms for the expected repulsive effect of the tank
wall and for the repulsive/attractive and alignment interactions between two
fish. These forms take into account the fish anisotropic perception of its
physical and social environment and must satisfy some specific symmetry
constraints which help us to differentiate these interactions and disentangle
their relative contributions. The amount and precision of data accumulated in
this work and this modeling approach allow us to reconstruct the actual
functional form of the response functions of fish governing their heading
changes as a function of the distance, orientation, and angular position relative
to an obstacle or a neighbor. We show that the implementation of these
interactions in a stochastic model of spontaneous burst-and-glide swimming quantitatively
reproduces the motion and spatial distributions observed in experiments with a
single fish and with two fish.

\section{Results}

\subsection{Characterization of individual swimming behavior}

\emph{Hemigrammus rhodostomus} fish have been monitored
swimming alone and freely in shallow water in three different  circular tanks
of radius $R=176$, 250, 353\,mm (see Supplementary Information (SI)  for details).
This species performs a burst-and-coast type of
swimming characterized by sequences of sudden increase in speed
followed by a mostly passive gliding period (see SI Movie S1). This
allows the analysis of a trajectory as a series of discrete
decisions in time. One can then identify the candidate stimuli
(\emph{e.g.} the distance, the orientation and velocity of a
neighboring fish, or the distance and orientation of an obstacle)
that have elicited a fish response and reconstruct the associated
stimulus-response function. Most changes in fish heading occur
exactly at the onset of the acceleration phase.  We label each of
these increases as a ``kick''.


Figs.~\ref{Figure1}A and \ref{Figure1}B show typical trajectories
of  \emph{H. rhodostomus} swimming alone or in groups of two
fish. After the data treatment (see SI  and Fig.~S1 and S2 there), it is possible to identify each kick (delimited by vertical lines in
Figs.~\ref{Figure1}C and \ref{Figure1}D), which we use to describe
fish trajectories as a group of straight lines between each of these
events.
While the average duration between kicks is close to 0.5\,s for experiments with one
or two fish (Fig.~\ref{Figure1}G), the mean length covered between
two successive kicks is slightly lower for two fish (Fig.~\ref{Figure1}H). The typical velocity of the fish in their active periods (see SI) is of order 140\,mm/s.

\subsection{Quantifying the effect of the interaction of a single fish with the
wall}
\begin{figure}[ht]
	\centering
	\includegraphics[width=\columnwidth]{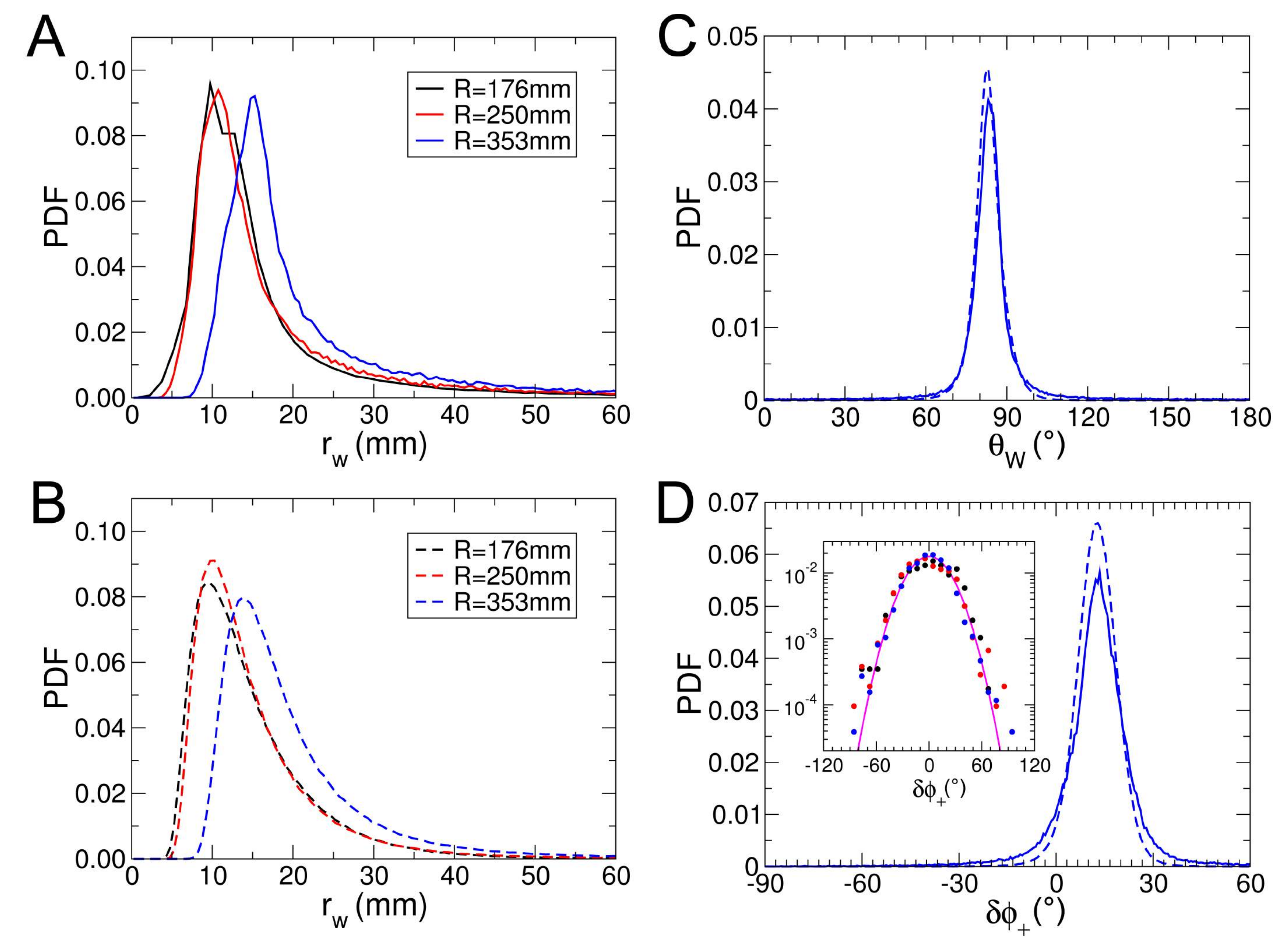}
	\caption{Quantification of the spatial distribution and motion of a fish
swimming alone. Experimental (A; full lines) and theoretical (B;
dashed lines) PDF of the distance to the wall $r_{\rm w}$ after a kick in the three arenas
of radius $R=176$, 250, 353\,mm. C: experimental (full line) and theoretical (dashed
line) PDF of the relative angle of the fish with the wall $\theta_{\rm w}$
($R=353\,$mm). D: PDF of the signed angle variation
$\delta\phi_+=\delta\phi{\times}{\rm Sign}(\theta_{\rm w})$ after each kick ($R=353\,$mm). The
inset shows the distribution of $\delta\phi_+$ when the fish is near the center of the tank
$(r_{\rm w}>60\,$mm), for $R=176$, 250, 353\,mm (colored dots), which becomes centered at
$\delta\phi_+=0^\circ$ and Gaussian of width $\approx 20^\circ$ (full line). \label{Figure2}}
\end{figure}

Fig.~\ref{Figure2}A shows the experimental probability density
function (PDF) of the distance to the wall $r_{\rm w}$ after each kick,
illustrating that the fish spends most of the time very close to the
wall. We will see that the combination of the burst-and-coast nature of the trajectories (segments of average length $\sim 70$\,mm, but smaller when the fish is very close to the wall) and of the narrow distribution of angle changes  between kicks (see Fig.~\ref{Figure2}D)  prevent a fish from efficiently escaping the curved wall of the tank (see SI Movie S2).
Fig.~\ref{Figure2}C shows the PDF of the
relative angle of the fish to the wall $\theta_{\rm w}$, centered near,
but clearly below 90$^\circ$, as the fish remains almost parallel to
the wall and most often goes toward it.

In order to characterize the behavior with respect to the walls, we
define the signed angle variation $\delta\phi_+=\delta\phi{\times}{\rm
Sign}(\theta_{\rm w})$ after each kick, where $\delta\phi$ is the measured
 angle variation.
$\delta\phi_+$ is hence positive when the fish goes away
from the wall and negative when the fish is heading towards it. The
PDF of $\delta\phi_+$  is  wider than a
Gaussian and is clearly centered at a positive $\delta\phi_+\approx
15^\circ$ (tank of radius $R=353\,$mm), illustrating that the fish works at avoiding the wall
(Fig.~\ref{Figure2}D).
When one restricts the data to instances where the fish is at a
distance $r_{\rm w}>60\,$mm from the wall, for which its influence becomes
negligible (see Fig.~\ref{Figure3}A and the discussion hereafter),
the PDF of $\delta\phi_+$ indeed becomes symmetric,
independent of the tank in which the fish swims, and takes a quasi Gaussian form of
width of order 20$^\circ$ (inset of
Fig.~\ref{Figure2}D).
The various quantities displayed in Fig.~\ref{Figure2} will
ultimately be used to calibrate and test the predictions of our
model.

\subsection{Modeling and direct measurement of fish interaction with the wall}
We first define a simple model for the spontaneous
burst-and-coast motion of a single fish without any wall
boundaries, and then introduce the fish-wall interaction, before considering
the interaction between two fish in the next subsection D.
The large amount of data accumulated (more than 300000
recorded kicks for 1 fish, and 200000 for 2 fish; see SI) permits us to not only precisely characterize the
interactions, but also to test the model by comparing its results to
various experimental quantities which would be very sensitive to a
change in model and/or parameters (\emph{e.g.} the full fish-wall
and fish-fish distance and angle distributions instead of simply
their mean).

\subsubsection{Swimming dynamics without any interaction}

\begin{figure}[ht]
	\centering
	\includegraphics[width=0.9\columnwidth]{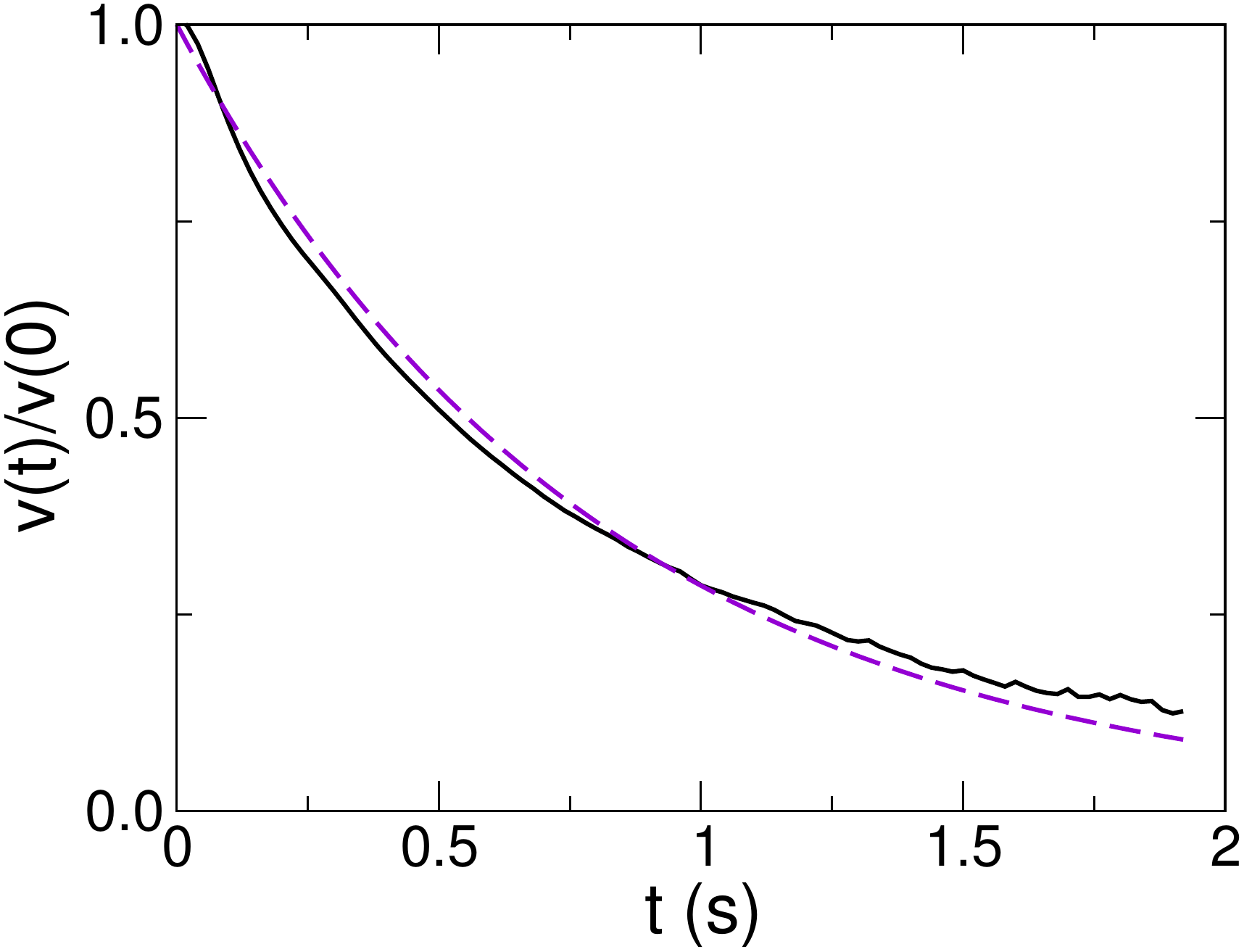}
	\caption{Average decay of the fish speed right after a kick (black line), which can be reasonably described by an exponential decay with a relaxation time  $\tau_0\approx 0.80\,$s (violet dashed line)\label{figv}}
\end{figure}

We model the burst-and-coast motion by a series of instantaneous
kicks each followed by a gliding period where fish travel in straight
lines with a decaying velocity. At the $n$-th kick, the fish located
at $\vec{x}_n$ at time $t_n$ with angular direction $\phi_n$
randomly selects a new heading angle $\phi_{n+1}$, a start or peak
speed $v_n$, a kick duration $\tau_n$, and a kick length $l_n$.
During the gliding phase, the speed is empirically found to decrease quasi
exponentially to a good approximation, as shown on  Fig.~\ref{figv},
with a decay or dissipation time $\tau_0\approx 0.80\,$s, so
that knowing $v_n$ and $\tau_n$ or $v_n$ and $l_n$, the third
quantity is given by
$l_n=v_n\tau_0(1-\exp[-\frac{\tau_n}{\tau_0}])$. At the end of the
kick, the position and time are updated to
\begin{equation}
\vec{x}_{n+1}=\vec{x}_n+l_n\vec{e}(\phi_{n+1}), \quad
t_{n+1}=t_n+\tau_n,
\end{equation}
where $\vec{e}(\phi_{n+1})$ is the unit vector along the new angular
direction $\phi_{n+1}$ of the fish.
In practice, we generate $v_n$ and $l_n$, and hence $\tau_n$ from
simple model bell-shaped probability density functions (PDF)
consistent with the experimental ones shown in Figs.~\ref{Figure1}G and \ref{Figure1}H. In addition, the distribution of
$\delta\phi_{\rm R}=\phi_{n+1}-\phi_{n}$ (the ${\rm R}$ subscript stands for
``random'') is experimentally found to be very close to a Gaussian distribution
when the fish is located close to the center of the tank, $i.e.$
when the interaction with the wall is negligible (see the inset of
Fig.~\ref{Figure2}D). $\delta\phi_{\rm R}$ describes the spontaneous
decisions of the fish to change its heading:
\begin{equation}
\phi_{n+1}=\phi_{n} +\delta\phi_{\rm R}=\phi_{n}+\gamma_{\rm R}\,g,
\label{phione}
\end{equation}
where $g$ is a Gaussian random variable with zero average and unit
variance, and $\gamma_{\rm R}$ is the intensity of the heading
direction fluctuation, which is found to be of order 0.35 radian
($\approx 20^\circ$) in the three tanks.

By exploiting the burst-and-coast dynamics of \emph{H.
rhodostomus}, we have defined an effective kick dynamics, of length
and duration $l_n$ and $\tau_n$. However, it can be useful to
generate the full continuous time dynamics from this discrete
dynamics. For instance, such a procedure is necessary to produce
``real-time'' movies  of fish trajectories obtained from the model (see SI Movies S2, S4, and S5).
As already mentioned, during a kick, the speed is
empirically found to decrease exponentially to a good approximation
(see Fig.~\ref{figv}), with a decay or dissipation time
$\tau_0\approx 0.80\,$s. Between the time  $t_{n}$ and $t_{n+1}=t_{n}+\tau_n$, the viscous
dynamics due to the water drag for $0\leq t\leq \tau_n$ leads to
\begin{equation}
\vec{x}(t_n+t)=\vec{x}_n+
l_n\frac{1-\exp[-\frac{t}{\tau_0}]}{1-\exp[-\frac{\tau_n}{\tau_0}]}\vec{e}(\phi_{n+1}),
\end{equation}
so that one recovers
$\vec{x}(t_n+\tau_n)=\vec{x}(t_{n+1})=\vec{x}_{n}+l_n\vec{e}(\phi_{n+1})=\vec{x}_{n+1}$.

\subsubsection{Fish interaction with the wall}

In order to include the interaction of the fish with the wall, we
introduce an extra contribution $\delta\phi_{\rm W}$
\begin{equation}
\delta\phi=\delta\phi_{\rm R}(r_{\rm w}) + \delta\phi_{\rm W}(r_{\rm w},\theta_{\rm w}),
\label{phionewall}
\end{equation}
where, due to symmetry constraints in a circular tank,
$\delta\phi_{\rm W}$ can only depend on the distance to the wall $r_{\rm w}$,
and on the angle $\theta_{\rm w}$ between the fish angular direction
$\phi$ and the normal to the wall (pointing from the tank center to
the wall; see Fig.~\ref{Figure1}E). We did not observe any
statistically relevant left/right asymmetry, which imposes the
symmetry condition
\begin{equation}
\delta\phi_{\rm W}(r_{\rm w},-\theta_{\rm w})=-\delta\phi_{\rm W}(r_{\rm w},\theta_{\rm w}).
\label{simw}
\end{equation}
The random fluctuations of the fish direction are expected to be
reduced when it stands near the wall, as the fish has less room for large
angles variations (compare the main plot and the inset of Fig.~\ref{Figure2}D), and we now define
\begin{equation}
\delta\phi_{\rm R}(r_{\rm w})=\gamma_{\rm R}[1-\alpha f_{\rm w}(r_{\rm w})]g.
\label{phionew}
\end{equation}
$f_{\rm w}(r_{\rm w})\to 0$, when $r_{\rm w}\gg l_{\rm w}$ (where $l_{\rm w}$ sets the range of
the wall interaction), recovering the free spontaneous motion in
this limit. In addition, we define $f_{\rm w}(0)=1$ so that the fluctuations
near the wall are reduced by a factor $1-\alpha$, which is found
experimentally to be close to $1/3$, so that $\alpha\approx 2/3$.

If the effective ``repulsive force'' exerted by the wall on the fish (first considered as a physical particle)
tends to make it go toward the center of the tank, it must take the
form $\delta\phi_{\rm W}(r_{\rm w},\theta_{\rm w})= \gamma_{\rm W}\sin(\theta_{\rm w})f_{\rm w}(r_{\rm w})$,
where the term $\sin(\theta_{\rm w})$ is simply the projection of the normal to
the wall (\textit{i.e.} the direction of the repulsion ``force'' due to the wall) on the angular acceleration of the fish (of direction
$\phi+90^\circ)$. For the sake of simplicity, $f_{\rm w}(r_{\rm w})$ is taken
as the same function as the one introduced in Eq.~(\ref{phionew}),
as it satisfies the same limit behaviors. In fact, a fish does not
have an isotropic perception of its environment. In order to take
into account this important effect in a phenomenological way, we
introduce
$\epsilon_{\rm w}(\theta_{\rm w})=\epsilon_{w,1}\cos(\theta_{\rm w})+\epsilon_{w,2}\cos
(2\theta_{\rm w})+...$, an even function (by symmetry) of $\theta_{\rm w}$,
which, we assume, does not depend on $r_{\rm w}$, and finally define
\begin{equation}
\delta\phi_{\rm W}(r_{\rm w},\theta_{\rm w})= \gamma_{\rm W}\sin(\theta_{\rm w})[1+\epsilon_{\rm w}(\theta_{\rm w})]f_{\rm w}(r_{\rm w}),
\label{phiwnew}
\end{equation}
where $\gamma_{\rm W}$ is the intensity of the wall repulsion.

Once the displacement $l$ and the total angle change $\delta\phi$
have been generated as explained above, we have to eliminate the
instances where the new position of the fish would be outside the tank. More
precisely, and since $\vec{x}$ refers to the position of the center
of mass of the fish (and not of its head) before the kick, we
introduce a ``comfort length'' $l_c$, which must be of the order of
one body length (BL; 1\,BL$\,\sim 30\,$mm; see SI), and we reject the move if the point
$\vec{x}+(l+l_c)\vec{e}(\phi+\delta\phi)$ is outside the tank. When
this happens, we regenerate $l$ and $\delta\phi$ (and in particular,
its random contribution $\delta\phi_{\rm R}$), until the new fish
position is inside the tank.
Note that in the rare cases where such a valid couple is not found
after a large number of iterations (say, 1000), we generate a new
value of $\delta\phi_{\rm R}$ uniformly drawn in $[-\pi,\pi]$ until a
valid solution is obtained. Such a large angle is for instance
necessary (and observed experimentally), when the fish happens to
approach the wall almost perpendicularly to it ($\delta\phi\sim
90^\circ$ or more; see Movie S5 at 20\,s, where the red fish performs such a large angle change).

In order to measure experimentally $\epsilon_{\rm w}(\theta_{\rm w})$ and
$f_{\rm w}(r_{\rm w})$, and confirm the functional form of Eq.~(\ref{phiwnew}),
we define a fitting procedure which is explicitly described in SI, by minimizing the error between the
experimental $\delta\phi$ and a general product functional form
$\delta\phi_{\rm W}(r_{\rm w},\theta_{\rm w})=f_{\rm w}(r_{\rm w})O_{\rm w}(\theta_{\rm w})$, where the only
constraint is that $O_{\rm w}(\theta_{\rm w})$ is an \textit{odd} function of
$\theta_{\rm w}$ (hence the name $O$), in order to satisfy the symmetry condition of Eq.~(\ref{simw}).
Since multiplying $O_{\rm w}$ by an
arbitrary constant and dividing $f_{\rm w}$ by the same constant leaves
the product unchanged, we normalize $O_{\rm w}$ (and all angular functions
appearing below) such that its average square is unity:
$\frac{1}{2\pi}\int_{-\pi}^{+\pi} O_{\rm w}^2(\theta_{\rm w})\,d\theta_{\rm w}=1$.

\begin{figure}[ht]
	\centering
	\includegraphics[width=\columnwidth]{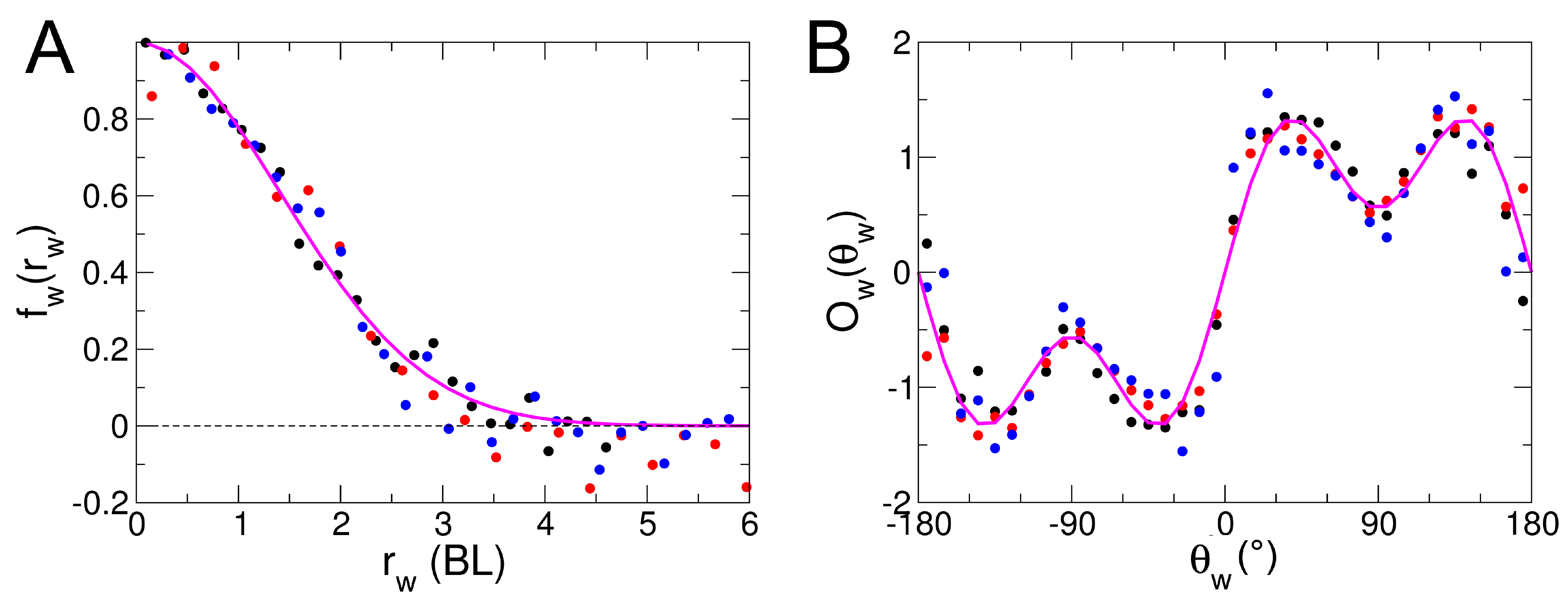}
	\caption{Interaction of a fish with the tank wall as a function of its distance $r_{\rm w}$ (A) and
its relative orientation  to the wall $\theta_{\rm w}$(B) as measured experimentally
in the three tanks of radius $R=176\,$mm (black), $R=250\,$mm (blue),
$R=353\,$mm (red). The full lines correspond to the analytic forms
of $f_{\rm w}(r_{\rm w})$ and $O_{\rm w}(\theta_{\rm w})$ given in the text. In particular,
$f_{\rm w}(r_{\rm w})$ is well approximated by a Gaussian of width $l_{\rm w}\approx
2\,$BL$\sim 60\,$mm. \label{Figure3}}
\end{figure}

For each of the three tanks, the result of this procedure is presented
as a scatter plot in Figs.~\ref{Figure3}A and \ref{Figure3}B
respectively, along with the simple following functional forms
(solid lines)
\begin{eqnarray}
O_{\rm w}(\theta_{\rm w})\propto \sin(\theta_{\rm w})[1+0.7\cos
(2\theta_{\rm w})],\label{gw}\\
f_{\rm w}(r_{\rm w}) =\exp\left[-\left(r_{\rm w}/l_{\rm w}\right)^2\right],~{\rm with }~ l_{\rm w}\approx
2\,{\rm BL}.\label{fw}
\end{eqnarray}
Hence, we find that the range of the wall interaction is of order
$l_{\rm w}\approx 2\,{\rm BL}\sim 60\,$mm, and is strongly reduced when
the fish is parallel to the wall (corresponding to a  ``comfort''
situation), illustrated by the deep (\emph{i.e.} lower response)
observed for $\theta_{\rm w}\approx 90^\circ$ in Fig.~\ref{Figure3}B
($\cos (2\theta_{\rm w})\approx -1$). Moreover, we do not find any significative dependence of these functional forms with the radius of the tank, although the interaction strength $\gamma_W$ is found to decrease as the radius of the wall increases (see Table S3). The smaller is the tank radius (of curvature), the more effort is needed by the fish to avoid the wall.

Note that the fitting procedure used to produce the results of
Fig.~\ref{Figure3} (described in detail in the SI) does not involve any regularization scheme imposing the scatter plots to fall on actual  continuous curves. The fact that they actually do describe such fairly smooth curves (as we will also find for the interaction functions between two fish; see Fig.~\ref{Figure5}) is an implicit va\-li\-da\-tion of our procedure.

In Fig.~\ref{Figure2}, and for the three tank radii considered, we compare the distribution of distance to the wall $r_w$, relative angle to the wall $\theta_w$, and angle change $\delta\phi$ after each kick, as obtained experimentally and in extensive numerical simulations of the model, finding an overall satisfactory agreement. On a more qualitative note, the model fish dynamics mimics fairly well the behavior and motion of a real fish (see SI Movie S2).

\subsection{Quantifying the effect of interactions between two fish}

Experiments with two fish were performed using the tank of radius $R=250\,$mm; and a total of around 200000 kicks were recorded (see SI for details).

\begin{figure}[ht]
	\centering
	\includegraphics[width=\columnwidth]{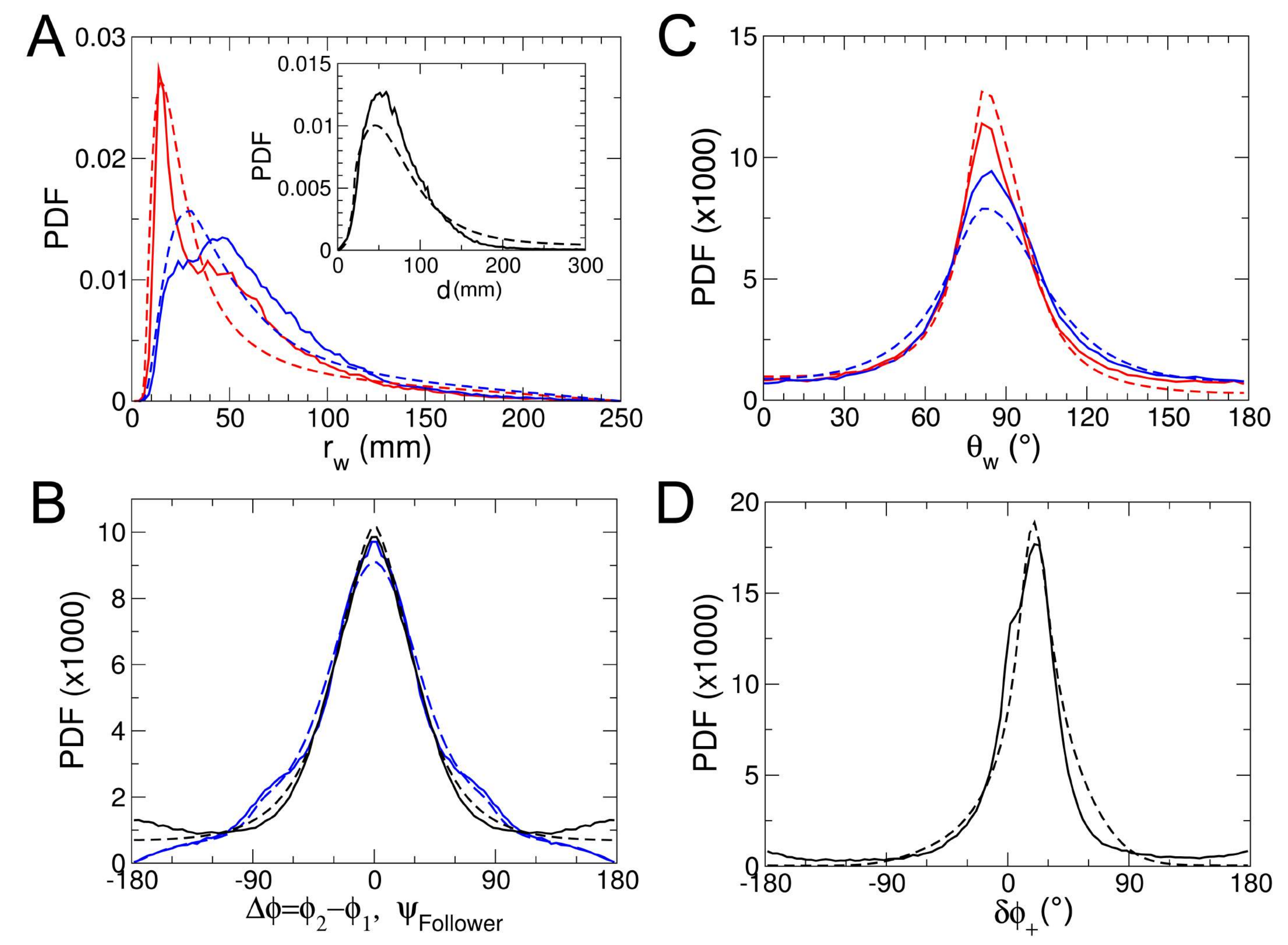}
	\caption{Quantification of the spatial distribution and motion
in groups of two fish. In all graphs, full lines correspond to
experimental results and dashed lines to numerical simulations of
the model. A: PDF of the distance to the wall, for the geometrical
leader (red) and follower (blue) fish; the inset displays the PDF
of the distance $d$ between the two fish. B: PDF of the relative
orientation $\Delta\phi=\phi_2-\phi_1$ between the two fish (black)
and PDF of the viewing angle $\psi$ of the follower (blue). C: PDF
of the relative angle to the wall $\theta_{\rm w}$ for the leader (red)
and follower fish (blue). D: PDF (averaged over both fish) of the
signed angle variation $\delta\phi_+=\delta\phi{\times}{\rm
Sign}(\theta_{\rm w})$ after each kick.
        \label{Figure4}}
\end{figure}

In Fig.~\ref{Figure4}, we present various experimental PDF which
characterize the swimming behavior of two fish resulting from their
interaction, and which will permit to
calibrate and test our model. Fig.~\ref{Figure4}A shows the PDF of
the distance to the wall, for the geometrical ``leader'' and
``follower'' fish. The geometrical leader is defined as the fish
with the largest viewing angle  $|\psi|\in[0,180^\circ]$ (see Fig.~\ref{Figure1}F
where the leader is the red fish), that is, the fish which needs to
turn the most to directly face the other fish. Note that the
geometrical leader is not always the same fish, as they can
exchange role (see SI Movie S5). We find that the geometrical leader
is much closer to the wall than the follower, as the follower tries
to catch up and hence hugs the bend. Still, both fish are farther
from the wall than an isolated fish is (see Fig.~\ref{Figure2}A; also compare SI Movie S2 and S4).
The inset of Fig.~\ref{Figure4}A shows the PDF of the distance
$d$ between the two fish, illustrating the strong attractive
interaction between them.

Fig.~\ref{Figure4}C shows the PDF of $\theta_{\rm w}$ for the leader and
follower fish, which are again much wider than for an isolated fish
(see Fig.~\ref{Figure2}C). The leader being closer and hence more
parallel to the wall displays a sharper distribution than the
follower. Fig.~\ref{Figure4}B shows the PDF of the relative
orientation $\Delta\phi=\phi_2-\phi_1$ between the two fish,
illustrating their tendency to align, along with the PDF of the
viewing angle $\psi$ of the follower. Both PDF are found to be very
similar and peaked at 0$^\circ$. Finally, Fig.~\ref{Figure4}D
shows the PDF (averaged over both fish) of the signed angle
variation $\delta\phi_+=\delta\phi{\times}{\rm Sign}(\theta_{\rm w})$ after each
kick, which is again much wider than for an isolated fish
(Fig.~\ref{Figure2}D). Due to their mutual influence, the fish swim farther from the wall than an isolated
fish, and the wall constrains less their angular fluctuations.

\subsection{Modeling and direct measurement of interactions between two fish}

In the presence of an other fish, the total heading angle change now reads
\begin{eqnarray}
\delta\phi&=&\delta \phi_{\rm R}(r_{\rm w}) +\delta\phi_{\rm W}(r_{\rm w},\theta_{\rm w})+\\
& &\delta \phi_{\rm Att}(d,\psi,\Delta\phi)+\delta\phi_{\rm Ali
} (d,\psi,\Delta\phi),\nonumber
\end{eqnarray}
where the random and wall contributions are given by
Eqs.~(\ref{phionew},\ref{phiwnew},\ref{gw},\ref{fw}), and the two
new contributions results from the expected attraction and alignment
interactions between fish. The distance between fish $d$, the
relative position or viewing angle $\psi$, and the relative
orientation angle $\Delta\phi$ are all defined in
Fig.~\ref{Figure1}F. By mirror symmetry already discussed in the
context of the interaction with the wall, one has the exact
constraint
\begin{equation}
\delta\phi_{\rm Att,\,Ali}(d,-\psi,-\Delta\phi)=-\delta\phi_{\rm Att,\,Ali}(d,\psi,
\Delta\phi),
\label{exsymint}
\end{equation}
meaning that a trajectory of the two fish observed from above the tank has the same probability of occurrence as the same trajectory as it appears when viewing it from the bottom of the tank.
We hence propose the following product expressions
\begin{eqnarray}
\delta\phi_{\rm Att}(d,\psi,\Delta\phi)=
F_{\rm Att}(d)O_{\rm Att}(\psi)E_{\rm Att}(\Delta\phi),\label{dphiatt}\\
\delta\phi_{\rm Ali}(d,\psi,\Delta\phi)=
F_{\rm Ali}(d)O_{\rm Ali}(\Delta\phi)E_{\rm Ali}(\psi),\label{dphiali}
\end{eqnarray}
where the functions $O$ are odd, and the functions $E$ are even. For
instance,  $O_{\rm Att}$ must be odd as the focal fish should turn by
the same angle (but of opposite sign) whether the other fish is at
the same angle $|\psi|$ to its left or right.
Like in the case of the wall interaction, we normalize the four angular functions appearing in Eqs.~(\ref{dphiatt},\ref{dphiali}) such that their average square is unity.
Both attraction and alignment interactions clearly break the law of action and reaction, as briefly mentioned in the Introduction and discussed in the Appendix. Although the heading angle difference perceived by the other fish is simply $\Delta\phi'=-\Delta\phi$, its viewing angle $\psi'$ is in general not equal to $-\psi$ (see Fig.~\ref{Figure1}F).

As already discussed
in the context of the wall interaction, an isotropic radial
attraction force between the two fish independent of the relative
orientation, would lead exactly to Eq.~(\ref{dphiatt}), with
$O_{\rm Att}(\psi)\sim \sin(\psi)$ and $E_{\rm Att}(\Delta\phi)=1$.
Moreover, an alignment force tending to maximize the scalar product,
$i.e.$ the alignment, between the two fish headings takes the
natural form $O_{\rm Ali}(\Delta\phi)\sim \sin(\Delta\phi)$, similar to
the one between two magnetic spins, for which one has
$E_{\rm Ali}(\psi)=1$. However, we allow here for more general forms
satisfying the required parity properties, due to the fish
anisotropic perception of its environment, and to the fact that its
behavior may also be affected by its relative orientation with the
other fish. For instance, we anticipate that $E_{\rm Ali}(\psi)$ should
be smaller when the other fish is behind the focal fish
($\psi=180^\circ$; bad perception of the other fish direction) than when it is ahead ($\psi=0^\circ$).

As for the dependence of $F_{\rm Att}$ with the distance between fish
$d$, we expect $F_{\rm Att}$ to be negative (repulsive interaction) at
short distance $d\leq d_0\sim 1\,$BL, and then to grow up to a
typical distance $l_{\rm Att}$, before ultimately decaying above
$l_{\rm Att}$. Note that if the attraction force is mostly mediated by
vision at large distance, it should be proportional to the 2D solid
angle produced by the other fish, which decays like $1/d$, for large
$d$. These considerations motivate us to introduce an explicit
functional form satisfying all these requirements
\begin{equation}
F_{\rm Att}(d)\propto\frac{d-d_0}{1+(d/l_{\rm Att})^2}.
\label{attraction}
\end{equation}

$F_{\rm Ali}$ should be dominant at short distance, before decaying for
$d$ greater than some $l_{\rm Ali}$ defining the range of the alignment
interaction. For large distance $d$, the alignment interaction
should be smaller than the attraction force, as it becomes more
difficult for the focal fish to estimate the precise relative
orientation of the other fish than to simply identify its presence.

\begin{figure}[ht]
	\centering
	\includegraphics[width=\columnwidth]{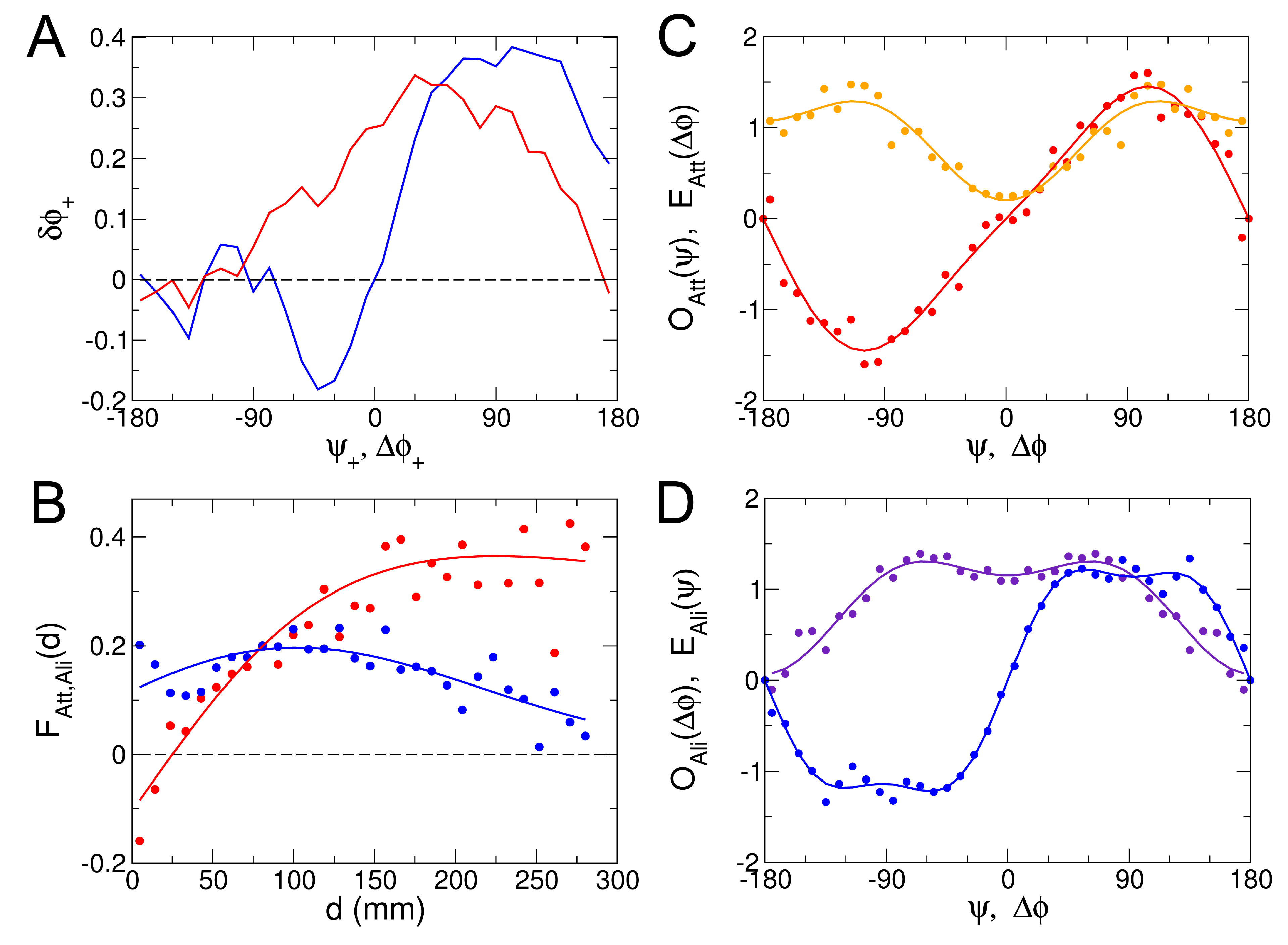}
	\caption{Quantification and modeling of interactions between
pairs of fish. A: we plot the average signed angle change after a
kick $\delta\phi_+=\delta\phi{\times}{\rm Sign}(\psi)$ $vs$
$\Delta\phi{\times}{\rm Sign}(\psi)$ (red) and
$\delta\phi_+=\delta\phi{\times}{\rm Sign}(\Delta\phi)$ $vs$ $\psi{\times}{\rm
Sign}(\Delta\phi)$ (blue) (see text). B: dependence of the
attraction ($F_{\rm Att}(d)$ in red) and alignment ($F_{\rm Ali}(d)$ in
blue) interactions with the distance $d$ between fish. The full lines
correspond to the physically motivated form of
Eq.~(\ref{attraction}) (red), and the fit proposed in the text for
$F_{\rm Ali}(d)$ (blue). C: $O_{\rm Att}(\psi)$ (odd function in
red) and $E_{\rm Att}(\Delta\phi)$ (even function in orange)
characterize the angular dependence of the attraction interaction,
and are defined in Eq.~(\ref{dphiatt}). D: $O_{\rm Ali}(\Delta\phi)$
(odd function in blue) and $E_{\rm Ali}(\psi)$ (even function in
violet), defined in Eq.~(\ref{dphiali}), characterize the angular
dependence of the alignment interaction. Dots in B, C, and D correspond to
the results of applying the procedure explained in SI
to extract the interaction functions from experimental data.\label{Figure5}}
\end{figure}

Fig.~\ref{Figure5}A shows  strong evidence for the existence of an
alignment interaction. Indeed, we plot the average signed angle
change after a kick $\delta\phi_+=\delta\phi{\times}{\rm Sign}(\psi)$ $vs$
$\Delta\phi{\times}{\rm Sign}(\psi)$ and $\delta\phi_+=\delta\phi{\times}{\rm
Sign}(\Delta\phi)$ $vs$ $\psi{\times}{\rm Sign}(\Delta\phi)$. In accordance
with Eqs.~(\ref{dphiatt},\ref{dphiali}), a strong positive
$\delta\phi_+$ when the corresponding variable is positive indicates
that the fish changes more its heading if it favors mutual alignment
(reducing $\Delta\phi$), for the same viewing angle $\psi$.

As precisely explained in SI, we have
determined the six functions appearing in
Eqs.~(\ref{dphiatt},\ref{dphiali}) by minimizing the error with the
measured $\delta\phi$, only considering kicks for which the focal
fish was at a distance $r_{\rm w}>2\,$BL from the wall, in order to
eliminate its effect (see Fig.~\ref{Figure3}A). This procedure
leads to smooth and well behaved measured functions displayed in
Fig.~\ref{Figure5}. As shown in Fig.~\ref{Figure5}B, the
functional form of Eq.~(\ref{attraction}) adequately describes
$F_{\rm Att}(d)$, with $l_{\rm Att}\approx 200\,$mm, and with an apparent
repulsive regime at very short range, with $d_0\approx 30\,{\rm mm}\sim 1\,$BL. The crossover between a
dominant alignment interaction to a dominant attraction interaction
is also clear. The blue full line in Fig.~\ref{Figure5}B, a guide
to the eye reproducing appropriately $F_{\rm Ali}(d)$, corresponds to
the phenomenological functional form
\begin{equation}
F_{\rm Ali}(d)\propto(d+d_0^\prime)\exp[-(d/l_{\rm Ali})^2],
\label{alignment}
\end{equation}
with $l_{\rm Ali}\approx 200\,$mm.
Note that  $F_{\rm Att}(d)$ and $F_{\rm Ali}(d)$ cannot be properly measured for $d>280$\,mm due to the lack of statistics, the two fish remaining most of the time close to each other (see the inset of Fig.~\ref{Figure5}A; the typical distance between fish is $d\sim 75\,$mm).

Fig.~\ref{Figure5}C shows
$O_{\rm Att}(\psi)\propto\sin(\psi)[1+\epsilon_{\rm Att,1}\cos(\psi)+...]$
(odd function) and $E_{\rm Att}(\Delta\phi)\propto
1+\eta_{\rm Att,1}\cos(\Delta\phi)+...$ (even function) along with fits
involving no more than 2 non zero Fourier coefficients (and often
only one; see SI  for their actual values).
$E_{\rm Att}(\Delta\phi)$ has a minimum for $\Delta\phi=0$ indicating
that the attraction interaction is  reduced when both fish are
aligned. Similarly, Fig.~\ref{Figure5}D shows
$O_{\rm Ali}(\Delta\phi)$ and $E_{\rm Ali}(\psi)$ and the corresponding fits.
As anticipated, the
alignment interaction is stronger when the influencing fish is ahead
of the focal fish ($|\psi|<90^\circ$), and almost vanishes when it is behind
($\psi=\pm 180^\circ$).

In Fig.~\ref{Figure4}, we compare the results of extensive
numerical simulations of the model including the interactions
between fish to experimental data, finding an overall
qualitative (SI Movie S4 and SI Movie S5) and quantitative
agreement.

As a conclusion of this section, we would like to discuss the generality of  the product functional forms of Eqs.~(\ref{dphiatt},\ref{dphiali}) for the interaction between fish,
or of Eq.~(\ref{phiwnew}) in the context of the wall interaction.
As already briefly mentioned, for a physical point
particle interacting through a physical force like gravity, the angle change
$\delta\phi_{\rm Att}(d,\psi)$ would be the projection of
the radial force onto the angular acceleration (normal to the
velocity of angular direction $\psi$ relative to the vector between the two particles) and would then exactly take the form $ F_{\rm Att}(d){\times}
\sin(\psi)$. Hence, Eq.~(\ref{dphiatt}) (resp. Eq.~(\ref{phiwnew}), for the wall interaction) is the simplest generalization accounting for the fish anisotropic perception of its environment, while keeping a product
form and still obeying the left/right symmetry condition of
Eq.~(\ref{exsymint}) (resp. of Eq.~(\ref{simw})). In principle,
$\delta\phi_{\rm Att}(d,\psi,\Delta\phi)$ should be written most
generally as an expansion $\sum_i
F_{{\rm Att},i}(d)O_{{\rm Att},i}(\psi)E_{{\rm Att},i}(\Delta\phi)$. However, as the
number of terms of this expansion increases, we run the risk of
overfitting the experimental data by the procedure detailed in the SI. In addition, the leading term of this expansion would
still capture the main behavioral effects of the interaction and
should be very similar to the results of Fig.~\ref{Figure5}, while the weaker
remaining terms would anyway be difficult to interpret. Note that the same argument applies to the
alignment interaction, when exploiting the analogy with the magnetic
alignment force between two spins. Eq.~(\ref{dphiali}) is the
simplest generalization of the interaction
$\delta\phi_{\rm Ali}(d,\Delta\phi)=F_{\rm Ali}(d)\sin(\Delta\phi)$ obtained
in this case, while preserving the left/righ symmetry and product
form. Considering the fact that no regularization or smoothing procedure was used in our data analysis (see SI), the quality (low noise, especially for angular functions) of
the results presented in Figs.~\ref{Figure3} and \ref{Figure5} strongly suggests that the
generalized product forms used here capture most of the features of
the actual experimental angle change.

\section{Discussion and conclusion}

Characterizing the social interactions between individuals as well
as their behavioral reactions to the physical environment is a
crucial step in our understanding of complex collective dynamics
observed in many group living species and their impact on individual
fitness \cite{Camazine2001,Krause2002}. In the present work, we have analyzed the
behavioral responses of a fish to the presence in its neighborhood
of an obstacle and to a conspecific fish. In particular, we used the discrete
decisions (kicks) of \emph{H. rhodostomus} to control its heading during
burst-and-coast swimming as a proxy to measure and model
individual-level interactions. The large amount of data accumulated
allowed us to disentangle and quantify the effects of
these interactions on fish behavior with a  high level of
accuracy.

We have quantified the spontaneous swimming behavior of a fish and modeled it by a kick dynamics with Gaussian distributed angle changes. We found that the interactions of fish with an obstacle and a
neighboring fish result from the combination of four behavioral
modes:

(1) wall avoidance, whose effect starts to be effective when
the fish is less than 2\,BL from a wall;

(2) short-range repulsion between fish,
when inter-individual distance is less than 30\,mm ($\sim 1$\,BL);

(3) attraction to the neighboring fish, which reaches a maximum value around
200\,mm ($\sim 6$ to 7\,BL) in our experimental conditions;

(4) alignment to the neighbor, which saturates around 100\,mm ($\sim
3$\,BL).

In contrast to previous phenomenological models, these behavioral modes are
not fixed to discrete and somewhat arbitrary zones of distances in
which the neighboring fish are found
\cite{Aoki1982,Huth1992,couzin2002}. Instead, there is a continuous
combination of attraction and alignment as a function of the
distance between fish. Alignment dominates attraction up to $\sim
75$\,mm ($\sim 2.5$\,BL) while attraction becomes dominant for
larger distances. As distance increases even more, attraction must
decrease as well. However, the limited size of the experimental
tanks and the lack of sufficient data for large $d$ prevented us from measuring this effect, suggesting the
long-range nature of the attraction interaction mediated by vision.
Note that a cluster of fish can elicit a higher level of attraction,
proportional to the 3D solid angle of the fish group as seen by the
focal fish, as suggested by models based on visual perception
\cite{Pita2015,Collignon2016}, and as captured by the power-law decay
proposed in Eq.~(\ref{attraction}). Designing experiments to test and quantify the long-range nature of the attraction interaction between fish would be of clear interest.

Moreover, the behavioral
responses are strongly modulated by the anisotropic perception of
fish. The wall repulsion effect is maximum when the orientation of
the fish with regards to the wall is close to 45$^\circ$ and minimum
when the fish is parallel to the wall. Likewise, the maximum
amplitude alignment occurs when a neighboring fish
is located on the front left or right and vanishes as its position
around the focal fish moves towards the back.

To quantify separately the effects of attraction and alignment,
we  exploited physical analogies and symmetry
considerations to extract the interactions between a focal fish and
the wall and with another fish. Previous studies have shown that in
the Golden shiners \cite{Katz2011} and the Mosquito fish
\cite{Herbert2011}, there was no clear evidence for an explicit
matching of body orientation. In these species, the alignment
between fish was supposed to results from a combination of
attraction and repulsion. However, at least in the Mosquito fish, it
is likely that the strength of alignment could have been
underestimated because the symmetry constraints on alignment and
attraction were not taken into consideration. In the Rummy-nose
tetra, we find strong evidence for the existence of an explicit
alignment.

The characterization and the measurement of burst-and-coast swimming and
individual interactions were then used to build and calibrate a model that
quantitatively reproduces the dynamics of swimming of fish alone and in groups
of two and the consequences of interactions on their spatial and angular
distributions. The model shows that the wall avoidance behavior coupled with the
burst-and-coast motion results in an unexpected concentration of fish
trajectories close to the wall, as observed in our experiments. In fact, this
phenomenon is well referenced experimentally for run-and-tumble swimming (for
instance, in sperm cells \cite{RTexp1} or bacteria \cite{RTexp2}). It can be
explained theoretically and reproduced in simple models \cite{RTth1,RTth2}, as
the effective discreteness of the trajectories separated in bursts or tumbles
prevents the individuals from escaping the wall. Our model also reproduces the
alternation of temporary leaders and followers in groups of two fish, the
behavior of the temporary leader being mostly governed by its interactions with the
wall, while the temporary follower is mostly influenced by the behavior of the
temporary leader.

This validated model can serve as a basis for testing hypotheses on the combination of influence exerted by multiples neighbors on a focal fish in tanks of arbitrary shape. Moreover, it would certainly be interesting to study theoretically the dynamics of many fish swimming without any boundary and according to the found interactions. The study of the phase diagram as a function of the strength of the attraction and alignment interactions (and possibly their range) should show the emergence of various collective phases (schooling phase, vortex phase...) \cite{Tunstrom2013,Calovi2014}.

Finally, our method has proved successful in disentangling and fully characterizing the interactions that govern the behavior of pairs of animals when large amount of data are available.
Hence, it could be successfully applied to collective motion phenomena occurring
in various biological systems at different scales of organization.

\section*{Contribution of authors}

C.S. and G.T. designed research;
D.S.C., V.L., U.L., and G.T. performed research; D.S.C., A.L., and C.S. developed the model;
D.S.C., A.L., V.L., U.L., H.C., C.S., and G.T.  analyzed data;
A.P.E. contributed new reagents/analytic tools; V.L., C.S., and G.T. wrote the paper.

\acknowledgments{This work was supported by grants from the Centre National
de la Recherche Scientifique and Universit\'e Paul Sabatier (project
Dynabanc). D.S.C. was funded by the Conselho Nacional de
Desenvolvimento Cientifico e Tecnol\'ogico -- Brazil. V.L. and U.L.
were supported by a doctoral fellowship from the scientific council
of the Universit\'e Paul Sabatier.}

\vskip 0.5cm
\appendix*
\section{Intelligent and dumb active matter}

A rather general equation describing the dynamics of a standard physical particle moving in a thermal bath (or a medium inducing a friction   and a random stochastic force, like a gas) and submitted to physical external forces $\vec{F}_{\rm Phys}(\vec{x})$ (due to other particles and/or external fields) reads
\begin{equation}
\frac{d\vec{v}}{dt}=-\frac{\vec{v}}{\tau}+\vec{F}_{\rm Phys}+\sqrt{\frac{2T}{\tau}}\vec{\eta},
\label{ap1}
\end{equation}
where $\vec{v}=\frac{d\vec{x}}{dt}$ is the particle velocity, $T$ is the temperature, and $\vec{\eta}(t)$ is a stochastic Gaussian noise, delta-correlated in time,
$\langle\vec{\eta}(t)\vec{\eta}(t')\rangle=\delta(t-t')$. In particular, if the physical force is conservative and hence is the gradient of a potential $V_{\rm Phys}(\vec{x})$, the stationary velocity and position probability distribution of the particle produced by this equation is well known to be the Boltzmann distribution,
\begin{equation}
P(\vec{x},\vec{v})=\frac{1}{Z}{\exp\left(-\frac{E}{T}\right)},
\label{ap2}
\end{equation}
where $E=\frac{v^2}{2}+V_{\rm Phys}$ is the energy, and $Z$ is a normalization constant.

\subsection{Dumb active matter}

An active particle is characterized by its intrinsic or desired velocity $\vec{u}$. Its actual velocity  $\vec{v}=\frac{d\vec{x}}{dt}$ rather generally obeys an equation similar to Eq.~(\ref{ap1}):
\begin{equation}
\frac{d\vec{v}}{dt}=-\frac{\vec{v}-\vec{u}}{\tau}+\vec{F}_{\rm Phys}+\sqrt{\frac{2T}{\tau}}\vec{\eta},
\label{ap3}
\end{equation}
where the first term on the right hand side tends to make the actual velocity go to the intrinsic velocity.
Eq.~(\ref{ap3}) has to be supplemented with a specific equation for the intrinsic velocity. Here, for the sake of simplicity, we assume that $\vec{u}$ is a simple Ornstein-Uhlenbeck stochastic process,
\begin{equation}
\frac{d\vec{u}}{dt}=-\frac{\vec{u}}{\tau'}+\sqrt{\frac{2T'}{\tau}}\vec{\eta'},
\label{ap4}
\end{equation}
where $\vec{\eta'}$ is an other stochastic Gaussian noise, uncorrelated with  $\vec{\eta}$, and $\tau'$ is some correlation time, a priori unrelated to $\tau$. In general, the stationary distribution $P(\vec{x},\vec{v},\vec{u})$ is not known, although some analytical results can be obtained in some limits (for instance, large friction, and separation of the time scales $\tau$ and $\tau'$) \cite{Col1,Col2}.

A first limiting case of this equation is the strong friction limit (small $\tau$), where the inertial term in Eq.~(\ref{ap3}) becomes negligible, leading to
\begin{equation}
\frac{d\vec{x}}{dt}=\vec{v}=\vec{u}+{\tau}\vec{F}_{\rm Phys}+\sqrt{2T\tau}\vec{\eta}.
\label{ap5}
\end{equation}
In the limit of a ``cold'' medium, where the stochastic force is absent or negligible, we obtain
\begin{equation}
\vec{v}=\vec{u}+{\tau}\vec{F}_{\rm Phys}.
\label{ap6}
\end{equation}
Note that in Eqs.~(\ref{ap3},\ref{ap4},\ref{ap5},\ref{ap6}), the physical force directly impacts the final velocity $\vec{v}$, but \emph{not} the intrinsic velocity $\vec{u}$ of the active particle. This very property constitutes our definition of a ``dumb'' active particle.

\subsection{Intelligent active matter}

As explained in the Introduction, animals can not only  be submitted to physical forces  $\vec{F}_{\rm Phys}$ (\emph{e.g.} a human physically pushing an other one), but mostly react to ``social forces'' $\vec{F}_{\rm Soc}$. These social interactions \emph{directly affect the intrinsic velocity} of the active particle, which constitutes our definition of an ``intelligent'' active particle. In the ``cold'' limit relevant for fish or humans (the substrate in which they move does not exert any noticeable random force), the system of equations  Eqs.~(\ref{ap3},\ref{ap4}) becomes
\begin{eqnarray}
\frac{d\vec{v}}{dt}=-\frac{\vec{v}-\vec{u}}{\tau}+\vec{F}_{\rm Phys},\label{ap7}\\
\frac{d\vec{u}}{dt}=-\frac{\vec{u}}{\tau'}+\vec{F}_{\rm Soc}+\sqrt{\frac{2T'}{\tau}}\vec{\eta'},
\label{ap8}
\end{eqnarray}
In the context of animal and intelligent active matter, the stochastic noise $\vec{\eta'}$ modelizes the spontaneous motion -- the ``free will'' -- of the animal (see  Eq.~(\ref{phione}), for \emph{Hemigrammus rhodostomus}).

Moreover, we also already mentioned that these social forces are in general non conservative and hence  strongly break the action-reaction law, as they generally depend not only on the positions of the particles, but also on their velocities (their relative direction $\Delta\phi$ and the viewing angle $\psi$ and $\theta_{\rm w}$ defined in Fig.~\ref{Figure1}). In the present work, we have for instance showed how the interaction of \emph{Hemigrammus rhodostomus} with a circular wall depend not only on the distance to the wall, but also on the viewing angle $\theta_{\rm w}$ between the fish heading and the normal to the wall (see Fig.~\ref{Figure3}). We also determined the dependence of the attraction and alignment interactions on the focal fish viewing angle $\psi$ and the two fish relative heading angle $\Delta\phi$ (see Fig.~\ref{Figure5}). Note that physical forces can induce a cognitive reaction and hence a change in the intrinsic velocity, so that  $\vec{F}_{\rm Soc}$ may also contain reaction term to the presence of physical forces $\vec{F}_{\rm Phys}$ (this was not the case in our experiments, except maybe, when the fish would actually touch the wall). Conversely, social interaction may lead to a particle willingly applying a physical force (a human moving toward an other one and then pushing her/him). As a consequence, the notion of a conserved energy and many other properties resulting from the conservative nature of standard physical forces are lost, leading to a much more difficult analytical analysis of the Fokker-Planck equation which can be derived from Eqs.~(\ref{ap7},\ref{ap8}).

It is obviously a huge challenge to characterize these social interactions in animal groups, in particular to better understand the collective phenomena emerging in various contexts \cite{Camazine2001,Giardina2008,Sumpter2010}. The system of equations Eqs.~(\ref{ap7},\ref{ap8}), for specific social interactions, also presents a formidable challenge, for instance to determine the stationary distribution $P(\vec{x},\vec{v},\vec{u})$. In the absence of physical forces, and in the limit of fast reaction (small $\tau$), leading to a perfect matching between the velocity and the intrinsic velocity, we obtain
\begin{eqnarray}
\vec{v}=\frac{d\vec{x}}{dt}=\vec{u},\label{ap9}\\
\frac{d\vec{u}}{dt}=-\frac{\vec{u}}{\tau'}+\vec{F}_{\rm Soc}+\sqrt{\frac{2T'}{\tau}}\vec{\eta'}.
\label{ap10}
\end{eqnarray}
Interestingly, this system is formally equivalent to Eq.~(\ref{ap1}) for a standard physical particle, although   Eq.~(\ref{ap10}) is formally an equation for the intrinsic velocity, equal to the actual velocity in the considered limit. Yet, the resulting stationary state $P(\vec{x},\vec{v}=\vec{u})$ is in general not known, because of the non conservative nature of the social interactions discussed above and in the Introduction.

\end{document}